\documentclass{article} \usepackage{fortschritte}
\usepackage{amssymb,amsfonts,latexsym,bbm}
\usepackage[dvips]{graphics}
\def\bea{\begin{eqnarray}}                            
\def\beq{\begin{equation}}                      
\def\calc {{\cal C}}
\def\calg {{\cal G}}
\def\calh {{\cal H}}
\def\calq {{\cal Q}}
\def\cft  {conformal field theory}
\def\cfts {conformal field theories}
\def\complex{{\mathbbm C}}
\def\eea{\end{eqnarray}}                               
\def\eeq{\end{equation}}                       
\def\g    {{\mathfrak g}}
\newcommand\hsp[1] {\mbox{\hspace{#1 em}}}
\def\id   {\mbox{\sl id}}
\newcommand\includeourbeautifulpicture[1] {{\begin{picture}(0,0)(0,0)
          \scalebox{.33}{\includegraphics{ahoop03-#1.eps}} \end{picture}}}
\newcommand\nxt[1] {\\\raisebox{.12em}{\rule{.35em}{.35em}}\hsp{.6}#1}
\def\su   {\mathfrak{su}}
\def\SU   {\mathfrak{su}}

\begin {document}                 

\def\titleline{
Algebras in tensor categories \newtitleline
and coset conformal field theories}
\def\email_speaker{
{\tt schweigert@math.uni-hamburg.de}}
\def\authors{
J.\ Fr\"ohlich,\,\1ad~~J.\ Fuchs,\,\2ad~~I.\ Runkel,\,\3ad~~C.\ 
Schweigert\,\,\4ad\sp}
\def\addresses{
\1ad  
Institut f\"ur Theoretische Physik,\,\, ETH Z\"urich,\,\, 
CH\,--\,8093 Z\"urich,\, Switzerland\\
\2ad   
Institutionen f\"or fysik,\, Karlstads Universitet,\, 
S\,--\,\,651\,88 Karlstad,\, Sweden\\ 
\3ad 
\,\,Institut f\"ur Physik,\,\, Humboldt-Universit\"at,\,\, 
D\,--\,12\,489 Berlin,\, Germany \\
\4ad
Fachbereich Mathematik, Universit\"at Hamburg,\, 
D\,--\,20\,146 Hamburg,\, Germany\\
}
\def\abstracttext{
The coset construction is the most important tool to construct
rational conformal field theories with
known chiral data. For some cosets at small level, so-called maverick cosets, 
the familiar analysis using selection and identification rules breaks down. 
Intriguingly, this phenomenon is linked to the existence of exceptional modular 
invariants. Recent progress in CFT, based on studying algebras in tensor 
categories, allows for a universal construction of the chiral data of coset 
theories which in particular also applies to maverick cosets.
}

 \begin{flushright}  {~} \\[-1cm] {\sf hep-th/0309269} \\[1mm]
 {\sf HU-EP-03/68} \\
 {\sf September 2003} \end{flushright}

\vspace{-5mm}

\large
\makefront

\section{Coset conformal field theories}

The coset construction is among the oldest \cite{baha} tools for obtaining
rational two-dimensional conformal field theories and has been very 
successful. It has been used to construct prominent classes of models, such 
as (super-)Virasoro minimal models and Kazama-Suzuki models. Still, it presents
a number of mysteries, even in the case of unitary conformal field theories,
to which we will restrict ourselves in this contribution. 

The coset construction is based
on the following data: A (finite-dimensional, complex, reductive) Lie
algebra $\g$ together with a choice $k$ of levels, i.e.\ a positive
integer for each simple ideal of $\g$, and a Lie subalgebra $\g'$
of $\g$. The embedding of $\g'$ into $\g$ determines the levels $k'$ of $\g'$. 
The aim of the coset construction is to obtain conformal field theories whose
chiral data -- like conformal weights, fusion rules, braiding and fusing 
matrices -- are completely known, and moreover can be expressed entirely in 
terms of the chiral data for $({\g},k)$ and $({\g}',k')$. This goal can 
indeed be achieved. However, as will become evident below, the way
this aim is reached is quite a bit more subtle than one might anticipate.

At first sight, understanding coset theories proceeds according to the following 
well-known pattern: The $\g/\g'$ coset theory has a description in terms of 
a gauged WZW sigma model \cite{gaku2kpsy} with target space the Lie group $G$
(the compact simply connected covering group associated to 
$\g$), in which the action of the subgroup $G'$ of $G$ is gauged. For 
constructing the space of states, this immediately suggests to start with 
positive energy representations of the (centrally extended) loop algebra 
$\hat{\g}$ based on $\g$ at level $k$, and require, \`a la Gupta-Bleuler, 
that the states of the coset theory are annihilated by the positive modes of 
the $\g'$-currents $J^{a'}$ (and some more operators):
  $$ J^{a'}_n v = 0 \quad\ \mbox{ for } \ \  n\,{>}\,0   \,. $$
This selects the highest weight spaces of $\hat{\g}'$; 
the candidate spaces of states of coset theories are thus the {\em branching
spaces\/} $\calh^\lambda_{\lambda'}$ that appear in the decomposition
  $$ \calh_\lambda^{\g} = \mbox{\large$\bigoplus_{\lambda'}^{}$}\,
  \calh^\lambda_{\lambda'} \otimes \calh_{\lambda'}^{\g'} $$
of irreducible highest weight $\g$-modules $\calh_\lambda$ into $\g'$-modules 
$\calh_{\lambda'}$.

Looking at simple examples reveals the following properties of these spaces:
\nxt Some branching spaces can be zero.
\nxt Some branching spaces can be isomorphic -- not just as graded vector spaces, 
     but even as modules over the coset chiral algebra (which is the commutant
     of the chiral algebra of $(\g',k')$ in the chiral algebra of $(\g,k)$).
\nxt Some branching spaces can be reducible as modules over the coset
     chiral algebra. 
\\[2pt]
The first two features -- selection rules and ``field identification'' -- 
already arise in the simplest example, the Ising model, which is realized by 
the coset
  $$ \frac{\SU(2)_1 \times \SU(2)_1}{\SU(2)_{2}} \, . $$
The branching spaces $\calh^{l_1,l_2}_{l}$ of this theory are labeled by three 
integers $l_{1,2}\,{\in}\,\{0,1\}$ and $l\,{\in}\,\{0,1,2\}$ (twice the 
respective $\su(2)$ spins). By the spin coupling rules, branching spaces 
can be non-zero only if $l_1\,{+}\,l_2\,{-}\,l$ is even.  Moreover, 
$\calh^{l_1,l_2}_{l}$ and $\calh^{1-l_1,1-l_2}_{2-l}$ are isomorphic Virasoro 
modules, reflecting the familiar symmetry of the Kac table of the Ising model.

The selection rules and field identifications are, at least at first
sight, well-understood in the Lagrangian setting: One actually gauges the 
{\em adjoint\/} action of the subgroup $G'$ on the group $G$ \cite{witt36}, 
so that the common center $Z\,{:=}\,Z(G)\,{\cap}\, Z(G')$ acts trivially. 
As a consequence, the group relevant for gauging is $G'\!{/}Z$, which is 
non-simply connected. Both the selection rules and the field identifications 
are implemented by summing over inequivalent $G'\!{/}Z$ bundles \cite{hori}.
In an algebraic formulation, simple currents are the 
appropriate concept to explain these effects \cite{scya5}; the selection rules
eliminate branching spaces of non-zero monodromy charge, and isomorphic 
branching spaces form simple current orbits. In this setting also the problem
of ``fixed point resolution'', i.e.\ of understanding the structure of 
reducible branching spaces, can be addressed \cite{fusS4}.

\section{Maverick coset theories}

In a {\em maverick coset theory\/} the pattern of field identifications and 
selection rules governed by simple currents breaks down -- there are more 
vanishing branching spaces, and more identifications.  The observation that 
such maverick cosets exist came as a big surprise. The first example was 
presented in \cite{dujo}; more examples were found in 
\cite{dujo2,fusS4,pesw}. A classification is not known to date, but in all 
known maverick cosets the level is small.

The existence of maverick cosets would not have been that astonishing, though,
had one only taken the lesson of {\em conformal embeddings\/} seriously. For a 
conformal embedding of $(\g',k')$ in $(\g,k)$ the Virasoro central charges of 
the respective WZW models coincide, $c_{(\g',k')}\,{=}\,c_{(\g,k)}$, so that 
the corresponding coset theory has central charge zero and hence is trivial
as a chiral CFT: For such cosets {\em all\/} branching spaces are either zero
or the trivial one-dimensional $c\,{=}\,0$ Virasoro module. Maverick cosets 
are thus intermediate between `ordinary' cosets and conformal embeddings.

Let us have a look at the simplest known example \cite{dujo}, $\su(2)$ 
embedded via its three-dimensional representation into $\su(3)$. The coset 
theory at level 2, $\su(3)_2 / \su(2)_8$, is ma\-ve\-rick. The ordinary 
selection rules allow all branching spaces $\calh^{(l_1\,l_2)}_q$ with $q$ 
even, and the expected identifications as Virasoro modules are 
$\calh^{(l_1\,l_2)}_q \,{\cong}\,\calh^{(l_1\,l_2)}_{8-q}$. However, 
comparison with the Kac table for the tetracritical Ising model 
shows that the branching spaces 
  $$ \calh_{2}^{(00)} \,{\cong}\, \calh_{6}^{(00)} ,\,\,
  \calh_{2}^{(20)} \,{\cong}\, \calh_{6}^{(20)} ,\,\,
  \calh_{2}^{(02)} \,{\cong}\, \calh_{6}^{(02)} ,\,\,
  \calh_{0}^{(10)} \,{\cong}\, \calh_{8}^{(10)} ,\,\,
  \calh_{0}^{(01)} \,{\cong}\, \calh_{8}^{(01)} ,\,\,
  \calh_{0}^{(11)} \,{\cong}\, \calh_{8}^{(11)}    $$
which a priori are allowed by the selection rules actually vanish as well,
and that there are additional identifications
\\[5pt]
  \begin{tabular}{llllll} $\qquad$
  &\quad$\calh_{0}^{(00)} \,{\cong}\, \calh_{8}^{(00)}
                          \,{\cong}\, \calh_{4}^{(11)}\hfill$ && $\chi(q)=\!\!$
    & $1\,~~~~~+\;q^2\, +2q^3 +3q^4 +4q^5 \,\cdots  $   \\[.15em]
  &$\quad\calh_{2}^{(11)} \,{\cong}\, \calh_{6}^{(11)}
                          \,{\cong}\, \calh_{4}^{(00)}\hfill$
  &&& $ 1\,{+}\,2q\,{+}\,2q^2\,{+}\,4q^3\,{+}\,5q^4 +8q^5 \,\cdots $   \\[.15em]
  &$\quad\calh_{2}^{(10)} \,{\cong}\, \calh_{6}^{(10)}
                          \,{\cong}\, \calh_{4}^{(02)}\hfill$
  &&& $1 +\,q +2q^2 +3q^3+5q^4 +7q^5 \,\cdots $   \\[.15em]
  &\quad$\calh_{2}^{(01)} \,{\cong}\, \calh_{6}^{(01)}
                          \,{\cong}\, \calh_{4}^{(20)}\hfill$
  &&& $1 +\,q +2q^2 +3q^3+5q^4 +7q^5 \,\cdots $    \\[.15em]
  &$\quad\calh_{0}^{(20)} \,{\cong}\, \calh_{8}^{(20)}
                          \,{\cong}\, \calh_{4}^{(01)}\hfill$
  &&& $1 +\,q +2q^2 +2q^3 +4q^4 +5q^5 \,\cdots $ \\[.15em]
  &$\quad\calh_{0}^{(02)} \,{\cong}\, \calh_{8}^{(02)}
                          \,{\cong}\, \calh_{4}^{(10)}\hfill$
  &&& $1 +\,q +2q^2 +2q^3 +4q^4 +5q^5 \,\cdots $ \\[.15em]
  \end{tabular}
\vskip 4pt

For all known maverick cosets, it has been observed that there exists
a modular invariant torus partition function of extension type for the
WZW based on $\g\,{\oplus}\,\g'$ in which
\nxt only such pairs $(\lambda,\lambda')$ of representations
     of $\g$ and $\g'$ appear that correspond to non-vanishing 
     branching spaces $\calh^\lambda_{\lambda'}$, and in which
\nxt the way these pairs are combined into irreducible representations of the
     extended chiral algebra reflects also the additional identifications. \\
Moreover, in all maverick cases this modular invariant for the 
$\g\,{\oplus}\,\g'$ WZW model is not of simple current type, but 
{\em exceptional\/}.
Having mentioned the term partition function, it is worth pointing out that
here our aim is to understand coset theories as {\em chiral} conformal field 
theories.\,%
  \footnote{~Indeed, already as chiral theories, they are useful in physical
  applications \cite{fpsw}.}
Accordingly, the $\g\,{\oplus}\,\g'$ torus partition function in question is 
nothing but the charge conjugation modular invariant with respect to the
extended chiral algebra.

Recently, the study of conformal field theories on surfaces with boundary
has given many new insights in the structure of modular invariant
partition functions (see \cite{fuRs+4} and references therein). In particular,
novel techniques for exceptional modular invariants have become 
available. In the rest of this note we present some of these techniques and 
explain how they allow us to gain a better understanding of maverick cosets.

\section{Algebras in tensor categories}

The first step is to find a convenient basis-independent way to encode
the chiral data of a given rational chiral CFT. This is provided by the 
representation category $\calc$ of the given chiral algebra, which has the 
structure of a {\em modular tensor category\/}. The objects 
$V$ of $\calc$ are representations of the chiral 
algebra, and the morphisms $f{:}\;V{\to}\,W$ of $\calc$ are intertwiners. 
Fusion of chiral algebra representations is encoded by a tensor product
  $$ \begin{array}l \otimes \,{:}\quad \calc \times \calc \,\to\, \calc
  \\[-.71em]{} \end{array}  $$


\noindent \parbox{29.8em}{%
that is associative, with the vacuum representation {\bf 1} acting as unit,
${\bf 1}\,{\otimes}\, V \,{=}\, V$. Braid group statistics, obeyed by quantum 
fields in low dimensions, yields a braiding on $\calc$, i.e.\ for every pair 
$(V,W)$ of objects an isomorphism $c_{V,W}{:}\; V\,{\otimes}\,W{\to}\,W\,
{\otimes}\,V$. Pictorially, the braiding is shown in figure 1.
Non-degeneracy of CFT two-point functions gives rise to a notion of dual object
(conjugate field), and the fractional part of the conformal weight
defines a `twist' for every object. 
               \\[.2pt]\mbox{$\ $}\hsp{0.68}
These structures are subject to quite
a few axioms, of course. Most of them just amount to the statement that
a visualization in terms of ribbon graphs like in figure 1 is possible.
In addition, it is required that the matrix with entries $s_{\lambda,\mu}
\,{:=}\, {\rm tr}(c_{\lambda,\mu}\,{\circ}\,c_{\mu,\lambda})$ is non-degenerate.
(This trace, depicted in figure 2, is the invariant of the Hopf link in
the three-manifold $S^3$; the $V_\mu$ are representatives for the isomorphism
classes of simple objects of $\calc$.)
               } \hfill
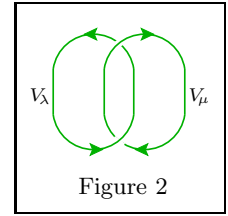
\begin{tabular}{r} {}\\[-11pt]
  \fbox{\parbox{3.7em}{
  \begin{picture}(26,84)(0,-26)
  \put(10,0)       {\includeourbeautifulpicture1}
  \put(7.2,-8.8)   {\scriptsize$V$}
  \put(7.8,49.6)   {\scriptsize$W$}
  \put(26.2,-8.8)  {\scriptsize$W$}
  \put(27.9,49.6)  {\scriptsize$V$}
  \put(4.5,-23)    {\small Figure~1}
                   \end{picture}    }}
  \\{}\\[-3pt]
  \fbox{\parbox{6.2em}{
  \begin{picture}(54,73)(0,-18)
  \put(10,0)       {\includeourbeautifulpicture2}
  \put(2.1,21.3)   {\scriptsize$V_{\!\lambda}$}
  \put(62.2,21.3)  {\scriptsize$V_{\!\mu}$}
  \put(20.5,-14)   {\small Figure~2}
                   \end{picture}    }}
\\[-1pt]{}\end{tabular}

In the case of coset theories we are given two modular tensor categories, 
$\calg$ and $\calg'$, for the chiral data of the WZW models based on $(\g,k)$ 
and $(\g',k')$. The goal we would like to achieve is then to express the 
category $\calq$ for the coset theory in terms of $\calg$ and $\calg'$.

To this end, we use the fact that algebra and representation theory
can be developed not just for (real or complex) vector spaces, but also in the 
much more general context of tensor categories. An algebra $(A,m,\eta)$ in a
tensor category $\calc$ consists of an object $A$ of $\calc$, a multiplication
$m{:}\; A\,{\otimes}\, A \,{\to}\, A$ that is associative, i.e.\ fulfils
  $$ m \circ (\id_A \,{\otimes}\, m) = m \circ (m \,{\otimes}\, \id_A) \,, $$
and a unit $\eta{:}\;{\bf1}\,{\to}\,A$. The latter should be thought of as the 
generalization of the map $\complex\,{\to}\,A$ that acts as $\xi\,{\mapsto}\,
\xi e$ for an algebra $A$ with unit element $e$ in the case that $A$ is a 
complex vector space. A (left-)\,representation $(M,\rho_M)$ of $A$ can be 
defined similarly: It is an object $M$ of $\calc$ together with a morphism
$\rho_M{:}\; A \,{\otimes}\, M \,{\to}\, M$ that obeys
  $$ \rho_M \circ ( m \,{\otimes}\, \id_M)
  = \rho_M \circ (\id_A \,{\otimes}\, \rho_M)
  \qquad{\rm and}\qquad
  \rho_M \circ (\eta \,{\otimes}\, \id_M) = \id_M \, .  $$

A particular class of algebras in modular tensor categories, called
{\em symmetric special Frobenius algebras\/}, is relevant in conformal field 
theory. All information about a full local CFT based on a given chiral CFT
is encoded in such an algebra \cite{fuRs+4}. For instance,
the coefficients of the torus partition function are given
by the dimension of the space of intertwiners of certain $A$-bimodules:
  $$ Z_{\lambda,\mu} = \dim{\rm Hom}_{A|A}(\alpha^-_{\!A}(V_\lambda),
  \alpha^+_{\!A}(V_\mu)) \, . $$

               \noindent
               \parbox{25.8em}{%
(It follows from general results that this is always modular invariant.)
This partition function is of extension type if $A$ is commutative
in the sense that $m\,{\circ}\, c_{A,A}\,{=}\, m$. In that case, the algebra 
describes just the vacuum sector of the extension, while the other sectors 
of the extended theory correspond to so-called {\em local\/} $A$-modules. 
Locality of an $A$-module means that the relation displayed in figure 3
is satisfied.
The category $\calc_{\!A}^{\scriptscriptstyle \ell{\rm oc}}$ of local modules 
over a symmetric special Frobenius algebra $A$ in a modular tensor category 
$\calc$ is again modular. 
This gives a very concise handle on the chiral data of the extended theory:
               } \hfill
  \fbox{\parbox{10.6em}
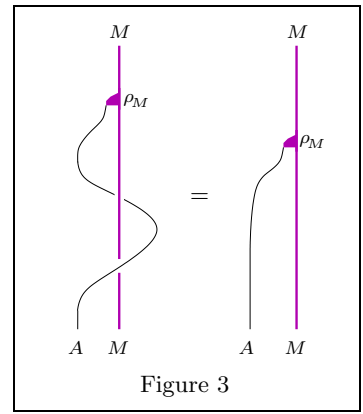
{
  \begin{picture}(54,147)(0,-27)
  \put(20.5,0)      {\includeourbeautifulpicture3}
  \put(17.2,-8.7)  {\scriptsize$A$}
  \put(32.1,-8.7)  {\scriptsize$M$}
  \put(32.7,111.2) {\scriptsize$M$}
  \put(38.2,86.9)  {\scriptsize$\rho_{\!M}^{}$}
  \put(63.5,49)    {\small$=$}
  \put(81.8,-8.7)  {\scriptsize$A$}
  \put(99.2,-8.7)  {\scriptsize$M$}
  \put(100.0,111.2){\scriptsize$M$} 
  \put(104.8,71.3) {\scriptsize$\rho_{\!M}^{}$}
  \put(44.5,-23)   {\small Figure~3}
                   \end{picture}    }}
  $$ \calc_{{\rm ext}} = \calc_{\!A}^{\scriptscriptstyle \ell{\rm oc}}
  \,. $$

The vacuum sector $\calh_\Omega^{\g}$ of the $\g$-theory decomposes in terms of 
sectors of the $\g'$-theory and of the coset theory according to
  $$ \calh_\Omega^{\g} = \mbox{\large$\bigoplus_{\lambda'}^{}$}\,
  \calh^\Omega_{\lambda'} \otimes \calh^{\g'}_{\lambda'}\, . $$
It follows that the modular tensor category $\calg$ of the $\g$-theory 
can be expressed as
  \beq  \calg = (\calq \,{\otimes}\, \calg')
  _{\!A}^{\scriptscriptstyle \ell{\rm oc}} \label G \eeq
through the categories $\calq$ and $\calg'$ and a suitable commutative 
symmetric special Frobenius algebra $A$ in $\calq \,{\otimes}\, \calg'$ that 
encodes the decomposition of the $\g$-vacuum $\calh_\Omega^{\g}$ given above. 

\smallskip

For the understanding of the coset category $\calq$ the following result 
is crucial:

\smallskip

\noindent
{\bf Theorem} \cite{ffrs}: 
There exists a (braided-) commutative symmetric special Frobenius algebra $B$ 
in the modular tensor category $\calg\,{\otimes}\, \overline{\calg'}$ such that
  \beq  \calq = (\calg\otimes \overline{\calg'})
  _{\!B}^{\scriptscriptstyle \ell{\rm oc}} \,. \label Q \eeq
This algebra $B$ corresponds to a modular invariant
of extension type for the $\g\,{\oplus}\,\g'$ theory.

\medskip

This is the desired expression of the chiral data $\calq$ of the coset
theory in terms of the chiral data $\calg$ and $\calg'$ of the parent WZW
models. The category $\overline{\calg'}$ in the theorem is obtained in a 
straightforward manner from $\calg'$. Basically, one applies complex 
conjugation to all chiral data. For details we refer to \cite{ffrs}. Also, $B$ 
can be constructed explicitly from the embedding $\g'\,{\hookrightarrow}\,\g$, 
and modularity of the tensor category $\calq$ is {\em derived\/} from the
modularity of $\calg$ and $\calg'$ (and from some much weaker assumptions on
$\calq$).

\medskip

A detailed discussion of the proof of the theorem is beyond the scope of this
note. Let us, however, mention three crucial aspects. 
\nxt 
In \cite{pare23brug2kios,ffrs}, commutative
algebra in a braided setting is developed. This theory turns out to be
much richer than ordinary commutative algebra for vector spaces (which is
already a rich theory). New notions, like local modules, two
different centers of an algebra, and new types of induction functors, play
an essential role in the proof. 
\nxt 
Another important ingredient is the unitarity of the modular matrix $S$ of 
rational CFTs. This property, which means that the braiding in two-dimensional 
CFTs is, in a sense, maximally non-degenerate, allows to ``solve'' the 
equality (\ref G) for $\calq$ in the form (\ref Q). 
\nxt 
All sectors appearing in the algebra $B$ contain fields of coset conformal 
weight $0$. The transition to local $B$-modules can therefore be thought of as
a means for removing ``redundant'' vacua in the collection of branching spaces.

\section{Conclusions}

Despite the progress reported here, the coset construction still
presents open problems:
\nxt The classification of coset theories $(\g,\g',k)$ that are maverick
     remains open.
\nxt Mavericks seem to be a low level phenomenon. For the group manifold as a
     sigma model target space, low level means large curvature. An
     understanding of the additional selection rules and identifications in
     maverick cosets as large curvature effects in a
     Lagrangian setting seems to be far beyond the reach of today's methods. 

Finally we point out that the results presented here arose within a 
larger research program \cite{fuRs+4,fuRs8}, which aims at constructing full 
local CFTs on world sheets of arbitrary topology from chiral CFTs using algebras 
in tensor categories. It is gratifying that the tools developed in this approach 
also shed new light on old and mysterious problems like the existence
of exceptional modular invariants associated to maverick coset theories. 
\\

\noindent
{\bf Acknowledgement.} 
This collaboration was supported in part by STINT 
(Stiftelsen f\"or internationalisering av h\"ogre utbildning och forskning,
grant IG 2001-070). I.R.\ is supported by the DFG project KL1070/2-1.
J.Fu., I.R.\ \&\ C.S.\ thank the ETH Z\"urich for hospitality;
C.S.\ thanks the organizers for the opportunity to present these results.
\newcommand\J[5]     {\ {\sl #5}, {#1} {#2} ({#3}) {#4}}
\newcommand\K[6]       {\ {\sl #6}, {#1} {#2} ({#3}) {#4}}
\newcommand\Prep[2]  {\ {\sl #2}, preprint {#1}}
\def\adma  {Adv.\wb Math.}
\def\comp  {Comm.\wb Math.\wb Phys.}
\def\ijmp  {Int.\wb J.\wb Mod.\wb Phys.\ A}
\def\joal  {J.\wB Al\-ge\-bra}
\def\josp  {J.\wb Stat.\wb Phys.}
\def\maan  {Math.\wb Annal.}
\def\mpla  {Mod.\wb Phys.\wb Lett.\ A}
\def\nupb  {Nucl.\wb Phys.\ B}
\def\phlb  {Phys.\wb Lett.\ B}
\def\phrd  {Phys.\wb Rev.\ D}
\def\wb{\,\linebreak[0]} 
\def\wB    {$\,$\wb}


\begin{thebibliography}{77}
\bibitem{baha}  {K.\ Bardak\c ci and M.B.\ Halpern,
        \J\phrd3{1971}{2493} {New dual quark models}}
\bibitem{gaku2kpsy} {K.\ Gaw\c edzki and A.\ Kupiainen,
        \J\phlb{215}{1988}{119} {G/H \cft\ from gauged WZW model};~
                     D.\ Karabali, Q-H.\ Park, H.J.\ Schnitzer, and Z.\ Yang,
        \J\phlb{216}{1989}{307} {A GKO construction based on a path
        integral formulation of gauged Wess-Zu\-mi\-no-Wit\-ten actions}}
\bibitem{witt36}{E.\ Witten,
        \J\nupb{371}{1992}{191} {The $N$ matrix model and gauged WZW models}}
\bibitem{hori}  {K.\ Hori, \K\comp{182}{1996}1
        {hep-th/9411134} {Global aspects of gauged WZW models}}
\bibitem{scya5} {A.N.\ Schellekens and S.\ Yankielowicz, \J\nupb{334}{1990}
        {67} {Field identification fixed points in the coset construction}} 
\bibitem{fusS4} {J.\ Fuchs, A.N.\ Schellekens, and C.\ Schweigert,
        \J\nupb{461}{1996}{371} {The resolution of field identification 
        fixed points in diagonal coset theories}}
\bibitem{dujo}  {D.C.\ Dunbar and K.G.\ Joshi, \K\ijmp8{1993}{4103}
        {hep-th/9210122} {Characters for coset conformal field theories}} 
\bibitem{dujo2} {D.C.\ Dunbar and K.G.\ Joshi, \K\mpla8{1993}{2803}
        {hep-th/9309093} {Maverick examples of coset conformal field theories}}
\bibitem{pesw}  {B.\ Pedrini, C.\ Schweigert, and J.\ Walcher,
        \J\phlb{466}{1999}{206} {New maverick coset theories}}
\bibitem{fpsw}  {J.\ Fr\"ohlich, B.\ Pedrini, C.\ Schweigert, and J.\ Walcher,
        \J\josp{103}{2001}{527} {Universality in quantum Hall systems:
        coset construction of incompressible states}} 
\bibitem{fuRs+4} {J.\ Fuchs, I.\ Runkel, and C.\ Schweigert, \K\nupb{624}{2002}
        {452} {hep-th/0110133} {Conformal correlation functions, Frobenius 
        algebras and triangulations};~
        \K\nupb{646}{2002}{353} {hep-th/0204148}
        {TFT construction of RCFT correlators I: Partition functions}}
\bibitem{ffrs} {J.\ Fr\"ohlich, J.\ Fuchs, I.\ Runkel, and C.\ Schweigert,
        \Prep{math.CT/0309465} {Correspondences of ribbon ca\-te\-go\-ries}}
\bibitem{pare23brug2kios} {B.\ Pareigis,
        \J\joal{171}{1995}{413} {On braiding and dyslexia};~\\A.\ Brugui\`eres,
        \J\maan{316}{2000}{215} {Cat\'egories pr\'emodulaires,
        modularisations et invariants des vari\'et\'es de dimension $3$};~\\
            A.A.\ Kirillov and V.\ Ostrik, \K\adma{171}{2002}{183}
        {math.QA/0101219} {On a $q$-analog of McKay correspondence and the 
        ADE classification of $\widehat{\mathfrak{sl}}(2)$ \cfts}} 
\bibitem{fuRs8} {J.\ Fuchs, I.\ Runkel, and C.\ Schweigert, \Prep{hep-th/0306164}
        {TFT construction of RCFT correlators II: Unoriented world sheets}}
\end{thebibliography}
\end{document}